# The molecular pathology of genioglossus in obstructive sleep apnea

Meng-Han Zhang[1,2], Yue-Hua Liu[2*]

[1]School of Stomatology affiliated to Medical College, Zhejiang University, Hangzhou, China

[2]Shanghai Key Laboratory of Craniomaxillofacial Development and Diseases, Shanghai Stomatological Hospital & School of Stomatology, Fudan University, Shanghai, China

* Correspondence: Yue-Hua Liu (liuyuehua@fudan.edu.cn)

Yue-Hua Liu, Department of Orthodontics, Shanghai Stomatological Hospital, Fudan University, Shanghai 200001, China. Tel: +86-21-55664116; E-mail: liuyuehua@fudan.edu.cn.

**Abstract**

Obstructive sleep apnea (OSA) is a sleep respiratory disease characterized by sleep snoring accompanied by apnea and daytime sleeplessness. It is a complex disease, with the multifactorial etiology, and the pathology is incompletely understood. Genioglossus (GG), the largest dilator of upper airway, whose fatigue is strongly correlated to onset of OSA. This brief review was to investigate the pathogenesis of OSA targeting on GG from different risk factors as gender, obesity, and aging, and the molecular mechanism of GG injury in OSA pathogenesis. We hope to find the targeted molecular mechanism on GG in OSA treatment.

Obstructive sleep apnea (OSA) is a sleep disordered breathing disease characterized by recurrent collapse of the upper airway during sleep. Epidemiologic data show that the overall prevalence of self-reported OSA was 6.4% in females and 13.8% in males, increasing with the BMI (from <20 to ⩾40 kg·m−2) in both sexes[1]. The prevalence of middle-aged people is significantly increased, about 17% of men and 9% of women. OSA can cause systemic diseases, and is an independent pathogenic factor for various diseases such as cardiovascular, cerebrovascular, metabolic, neurological, and psychiatric systems[2-3].

The major feature of OSA pathogenesis is a perturbation between upper airway obstruction and neuro-muscular responses during sleep[4]. Chronic intermittent hypoxia (CIH) result from recurrent upper airway collapse, as well as repeated high-level anti-resistance activity, would make the upper airway dilator more susceptible to develop fatigue[5]. A wide range of treatment options are applied for OSA, including continuous positive airway pressure (CPAP), weight loss, oral appliances (OA), and surgery[6]. However, the treatment of OSA targets the narrow anatomic structure of the upper airway, but lacks an effective therapy for upper airway muscle dysfunction.

Genioglossus (GG), the largest dilator of upper airway, whose fatigue is strongly correlated to onset of apnea[7]. GG originates from the medial superior mental spine of the mandible, with the inferior fibers attached to the hyoid bone, the middle and superior fibers distributed in the tongue body[8]. The molecular mechanism of OSA in terms of the main signaling pathways and epigenetics alterations are introduced, such as microRNA, long non-coding RNA, and DNA methylation, as well as small molecular compounds as potential targets for OSA[9]. However, the molecular mechanism GG in OSA/CIH models remains unclear. The research group is committed to studying the pathological mechanism of OSA targeting on GG.

**1. Risk factors for adult OSA pathogenesis and therapy targeting on GG**

The main risk factors for adult OSA include not only anatomical stenosis, but also include gender, obesity, and aging[10]. Therefore, we now review the pathogenic mechanism of OSA regarding on these latter risk factors.

**1.1 Estrogen related physiological process influence GG fatigue from gender**

**difference**

There is strong evidence for an increased susceptibility of males (or decreased susceptibility of females) to OSA[11]. The research group observed that the electromyogram (EMG) activity of GG was affected by hormone levels in young female[12]. Then we performed ovariectomy in rats, and found reduced EMG activity of GG with decreased myosin heavy chain IIA (MyHCIIA) isoform distribution[13]. Estrogen (estradiol) could stimulate the expression of estrogen receptor α (ERα) and greatly improve activities of GG; While androgen (testosterone) inhibited the expression of androgen receptor (AR) and Erβ, and impaired GG fatigue resistance capacity[14]. Furthermore, testosterone treatment decreased the fatigue resistance in aged GG muscles by acting at sarcoplasmic reticulum $Ca^{2+}$-ATPase (SERCA) activity and SERCA gene expression[15]. This implied that hormone levels may affect the function of GG by affecting ERs and SERCA.

It is speculated that GG fatigue in OSA patients may be influenced by estrogen related physiological process. Then the research group established CIH rat models and found reduced contractile and duration properties of GG in CIH rats. CIH decreased the proportion of MyHCIIA, reduce the activity and expression of SERCA and mitochondrial cytochrome c oxidase in GG, then finally destroyed GG ultrastructures. Ovariectomy in rats exacerbated the above effect, while estrogen can partially reverse the effect of CIH in ovariectomized rats in part via regulation of the expression of ERα.[16-20]

Based on the preliminary understanding of estrogen's protective mechanism on GG function, the research group try to use estrogen substitutes to treat upper airway fatigue. Phytoestrogen genistein, coumestrol or resveratrol increased muscle fatigue resistance respectively, in part by up-regulation of ERs expression in ovariectomized rats[21-22]. Resveratrol dimer exhibited better protection than genistein and resveratrol, and expressed higher binding affinity for ERβ than for Erα[23].

**1.2 Obesity/Aging induced-GG injury aggravate OSA pathogenesis**

Obesity and aging are another two important risk factors for OSA[24-25]. The research group discovered that obesity induced-GG injury with mitochondrial dysfunction and excessive oxidative stress, and activated the mitochondrial-related apoptotic pathway[26]. Hypoxia and aging interact to form a vicious circle with upregulation of p53 and p21, which worsened GG injury.[27]

## 2. Molecular mechanism of GG injury in OSA pathogenesis

### 2.1 hypoxia inducible factors-1α (HIF-1α)

HIF-1α also plays an important role in GG fatigue in OSA pathogenesis. CIH induced the expression of HIF-1α in the GG and altered the physical properties towards a more fatigable phenotype, whereas estrogen inhibited the over-expression of HIF-1α[28]. The research group further revealed that increased HIF-1α expression mediated muscle fatigue with mitochondrial ultrastructure damage via AMPK/PGC-1β signalling pathway[29]. The molecular regulation of myoblasts under hypoxia were described, in which HIF-1α mediates Notch and Wnt signaling pathways[30]. Then the research group established skeletal muscle conditional HIF-1α KO mice, and found there was a slight decrease in muscle fiber type I, IIa, IIx fiber types but an increase in type IIb[31]. A transcriptomic analysis of GG from conditional HIF-1α KO mice showed multiple genes expression involved in the myogenesis, muscle development, and carbo hydrate metabolism[32].

### 2.2 Reactive oxygen species (ROS)

Reactive oxygen species (ROS) act as a medium in CIH-induced GG injury. ROS generation, MDA, and DNA damage were increased when GG myoblast was under hypoxic condition, lead to myoblast apoptosis[33]. The research group also found myoblast pyroptosis during OSA was results from excessive ROS via the NF-κB/HIF-1α signaling pathway[34]. Nowadays, we conducted an RNA-Seq analysis of GG in CIH mice, and found a total of 637 differentially expressed genes (DEGs)[35]. Our previews study confirmed that CIH exposure could increase the expression of phosphofructokinase muscle-specific isoform and change metabolism in CIH rats[36]. Regarding the enriched biological processes terms, more than 40% of DEGs were associated with terms related to the responses to endogenous stimuli, cellular component organization and metabolic processes. It was verified in vivo and in vitro that CIH can induce excess ROS and reduce collagen synthesis, leading to the decline in EMG$^{GG}$ function in CIH mice[35].

## 3. Prospect

The research group have been dedicated to the pathological mechanism and treatment of adult OSA and pediatric OSA for decades. Recently, the possibility of stem cell therapy was tested, and it is found that dental pulp stem cells secretome alleviated hypoxia-induced GG myoblasts injury via Wnt/β-catenin and AMPK/PGC-1α signaling pathway[53-54]. Our prospect is to targeted molecular mechanism on GG, so as to effectively improve GG fatigue resistance to upper airway muscle collapse.

**Declaration of interests**

The authors declare no competing interests.

## References

1. Huang T, Lin BM, Markt SC, Stampfer MJ, Laden F, Hu FB, Tworoger SS, Redline S. Sex differences in the associations of obstructive sleep apnoea with epidemiological factors. *Eur Respir J*. 2018 Mar 15;51(3):1702421. doi: 10.1183/13993003.02421-2017.
2. Jin J. Screening for obstructive sleep apnea. *JAMA*. 2022 Nov 15;328(19):1988. doi: 10.1001/jama.2022.20142.
3. Marin JM, Carrizo SJ, Vicente E, Agusti AG. Long-term cardiovascular outcomes in men with obstructive sleep apnoea-hypopnoea with or without treatment with continuous positive airway pressure: an observational study. *Lancet*. 2005 Mar 19-25;365(9464):1046-53. doi: 10.1016/S0140-6736(05)71141-7.
4. Saboisky JP, Butler JE, McKenzie DK, Gorman RB, Trinder JA, White DP, Gandevia SC. Neural drive to human genioglossus in obstructive sleep apnoea. *J Physiol*. 2007 Nov 15;585(Pt 1):135-46. doi: 10.1113/jphysiol.2007.139584.
5. Boyd JH, Petrof BJ, Hamid Q, Fraser R, Kimoff RJ. Upper airway muscle inflammation and denervation changes in obstructive sleep apnea. *Am J Respir Crit Care Med*. 2004 Sep 1;170(5):541-6. doi: 10.1164/rccm.200308-1100OC.
6. Zhang M, Liu Y, Liu Y, Yu F, Yan S, Chen L, Lv C, Lu H. Effectiveness of oral appliances versus continuous positive airway pressure in treatment of OSA patients: An updated meta-analysis. *Cranio*. 2019 Nov;37(6):347-364. doi: 10.1080/08869634.2018.1475278.
7. GStrollo PJ Jr, Soose RJ, Maurer JT, de Vries N, Cornelius J, Froymovich O, Hanson RD, Padhya TA, Steward DL, Gillespie MB, Woodson BT, Van de Heyning PH, Goetting MG, Vanderveken OM, Feldman N, Knaack L, Strohl KP; STAR Trial Group. Upper-airway stimulation for obstructive sleep apnea. *N Engl J Med*. 2014 Jan 9;370(2):139-49. doi: 10.1056/NEJMoa1308659.
8. McCausland T, Bordoni B. Anatomy, head and neck: genioglossus muscle. 2023 Jun 5. In: StatPearls [Internet]. Treasure Island (FL): StatPearls Publishing; 2023 Jan–.

9. Zhang M, Lu Y, Sheng L, Han X, Yu L, Zhang W, Liu S, Liu Y. Advances in molecular pathology of obstructive sleep apnea. *Molecules*. 2022 Dec 1;27(23):8422. doi: 10.3390/molecules27238422.
10. Qian Y, Dharmage SC, Hamilton GS, Lodge CJ, Lowe AJ, Zhang J, Bowatte G, Perret JL, Senaratna CV. Longitudinal risk factors for obstructive sleep apnea: A systematic review. *Sleep Med Rev*. 2023 Oct;71:101838. doi: 10.1016/j.smrv.2023.101838.
11. Jordan AS, McEvoy RD. Gender differences in sleep apnea: epidemiology, clinical presentation and pathogenic mechanisms. Sleep Med Rev. 2003 Oct;7(5):377-89. doi: 10.1053/smrv.2002.0260.
12. Wang J, Liu YH, Li Q, Hou YX. The influence of hormonal level to genioglossus muscle activity in normal young women. *Journal of Tongji University (Medical Science)*. 2006;(01):45-48. doi:10.3969/j.issn.1008-0392.2006.01.013
13. Liu YH, Jia SS, Hou YX. Effects of ovariectomy on rat genioglossal muscle contractile properties and fiber-type distribution. *Angle Orthod.* 2009;79(3):509-514. doi: 10.2319/031608-149.1
14. Qi J, Liu YH, Wang F, Shao X, Song WH. Effects of sex hormones on genioglossal muscle activities, estrogen and androgen receptor expression in adult rat. *Zhonghua Kou Qiang Yi Xue Za Zhi*. 2007;42(2):85-89. doi:10.3760/j.issn:1002-0098.2007.02.007
15. Liu YH, Qi J, Hou YX, Wang F. Effects of sex hormones on genioglossal muscle contractility and SR $Ca^{2+}$-ATPase activity in aged rat. *Arch Oral Biol.* 2008;53(4):353-360. doi: 10.1016/j.archoralbio.2007.10.009
16. Liu YH, Qi J, Jia SS. Effects of estrogen on activity and subunits expression of cytochrome c oxidase in mitochondria from genioglossus of rats subjected to chronic intermittent hypoxia. *Zhonghua Kou Qiang Yi Xue Za Zhi*. 2008;43(10):604-607. doi:10.3321/j.issn:1002-0098.2008.10.008
17. Hou YX, Jia SS, Liu YH. 17Beta-estradiol accentuates contractility of rat genioglossal muscle via regulation of estrogen receptor alpha. *Arch Oral Biol.* 2010;55(4):309-317. doi: 10.1016/j.archoralbio.2010.02.002
18. Liu YH, Huang Y, Shao X. Effects of estrogen on genioglossal muscle contractile properties and fiber-type distribution in chronic intermittent hypoxia Rats. *Eur J Oral Sci.* 2009;117(6):685-690. doi: 10.1111/j.1600-0722.2009.00681.x
19. Liu YH, Li W, Song WH. Effects of oestrogen on sarcoplasmic reticulum $Ca^{2+}$-ATPase activity and gene expression in genioglossus in chronic intermittent hypoxia rat. *Arch Oral Biol.* 2009;54(4):322-328. doi: 10.1016/j.archoralbio.2009.01.009
20. Qi J, Liu YH, Song WH, Shao X. Effects of estrogen on genioglossus contractile properties and cell ultrastructures in chronic intermittent hypoxia rats. *Journal of Practical Stomatology*. 2008;45(10):627-630. doi:10.3969/j.issn.1001-3733.2008.02.004
21. Huang Y, Liu YH. Effects of phytoestrogens on genioglossus contractile properties in ovariectomized rats exposed to chronic intermittent hypoxia may be independent of their estrogenicity. *Eur J Oral Sci.* 2011;119(2):128-135. doi: 10.1111/j.1600-0722.2011.00815.x
22. Li W, Liu YH. Effects of Phytoestrogen genistein on genioglossus function and oestrogen receptors expression in ovariectomized rats. *Arch Oral Biol.* 2009;54(11):1029-1034. doi: 10.1016/j.archoralbio.2009.09.002
23. Lu Y, Liu Y, Li Y. Comparison of natural estrogens and synthetic derivative on genioglossus function and estrogen receptors expression in rats with chronic intermittent hypoxia. *J Steroid Biochem Mol Biol*. 2014;140:71-79. doi: 10.1016/j.jsbmb.2013.12.006
24. Sung CM, Tan SN, Shin MH, Lee J, Kim HC, Lim SC, Yang HC. The site of airway collapse in sleep apnea, its associations with disease severity and obesity, and implications for mechanical

interventions. *Am J Respir Crit Care Med*. 2021 Jul 1;204(1):103-106. doi: 10.1164/rccm.202011-4266LE.

25. Gaspar LS, Álvaro AR, Moita J, Cavadas C. Obstructive sleep apnea and hallmarks of aging. *Trends Mol Med*. 2017 Aug;23(8):675-692. doi: 10.1016/j.molmed.

26. Chen Q, Han X, Chen M, Zhao B, Sun B, Sun L, Zhang W, Yu L, Liu Y. High-fat diet-induced mitochondrial dysfunction promotes genioglossus injury - a potential mechanism for obstructive sleep apnea with obesity. *Nat Sci Sleep*. 2021 Dec 23;13:2203-2219. doi: 10.2147/NSS.S343721.

27. Zhu LY, Yu LM, Zhang WH, Deng JJ, Liu SF, Huang W, Zhang MH, Lu YQ, Han XX, Liu YH. Aging induced p53/p21 in genioglossus muscle stem cells and enhanced upper airway injury. *Stem Cells Int*. 2020;2020:8412598. doi: 10.1155/2020/8412598

28. Jia SS, Liu YH. Down-regulation of hypoxia inducible factor-1 alpha: a possible explanation for the protective effects of estrogen on genioglossus fatigue resistance. *Eur J Oral Sci*. 2010;118(2):139-144. doi: 10.1111/j.1600-0722.2010.00712.x

29. Lu Y, Mao J, Han X, Zhang W, Li Y, Liu Y, Li Q. Downregulated hypoxia-inducible factor 1α improves myoblast differentiation under hypoxic condition in mouse genioglossus. *Mol Cell Biochem*. 2021 Mar;476(3):1351-1364. doi: 10.1007/s11010-020-03995-1.

30. Zhang WH, Yu LM, Han XX, Liu YH. Research advances on the effects of hypoxia on proliferation and differentiation of skeletal muscle cells and related mechanisms. *Progress in Physiological Sciences*. 2019, 50(5):7. doi:CNKI:SUN:SLKZ.0.2019-05-002.

31. Xu HY, Lu Y, Li YY, Liu YH. Effect of HIF-1α on muscle fiber differentiation in mice. *Journal of Tongji University (Medical Science)*. 2017;38(02):1-4. doi: 10.16118/j.1008-0392.2017.02.001

32. Hao T, Liu YH, Li YY, Lu Y, Xu HY. A transcriptomic analysis of physiological significance of hypoxia-inducible factor-1alpha in myogenesis and carbohydrate metabolism of genioglossus in mice. *Chin Med J (Engl)*. 2017;130(13):1570-1577. doi: 10.4103/0366-6999.208235

33. Ding W, Chen X, Li W, Fu Z, Shi J. Genistein protects genioglossus myoblast against hypoxia-induced injury through PI3K-Akt and ERK MAPK pathways. *Sci Rep*. 2017;7(1):5085. doi: 10.1038/s41598-017-03484-4

34. Yu LM, Zhang WH, Han XX, Li YY, Lu Y, Pan J, Mao JQ, Zhu LY, Deng JJ, Huang W, Liu YH. Hypoxia-induced ROS contribute to myoblast pyroptosis during obstructive sleep apnea via the NF-kappaB/HIF-1α signaling pathway. *Oxid Med Cell Longev*. 2019;2019:4596368. doi: 10.1155/2019/4596368

35. Zhang MH, Han XX, Lu Y, Deng JJ, Zhang WH, Mao JQ, Mi J, Ding WH, Wu MJ, Yu LM, Liu YH. Chronic intermittent hypoxia impaired collagen synthesis in mouse genioglossus via ROS accumulation: A transcriptomic analysis. *Respir Physiol Neurobiol*. 2023 Feb;308:103980. doi: 10.1016/j.resp.2022.103980.

36. Jia SS, Liu YH. Effects of estrogen on the expression of phosphofructokinase muscle-specific isoform in genioglossus of chronic intermittent hypoxia rats. *Zhonghua Kou Qiang Yi Xue Za Zhi*. 2010;45(10):627-630. doi:10.3760/cma.j.issn.1002-0098.2010.10.015

37. Zhang WH, Yu LM, Han XX, Pan J, Deng JJ, Zhu LY, Lu Y, Huang W, Liu SF, Li Q, Liu YH. The secretome of human dental pulp stem cells protects myoblasts from hypoxia induced injury via the Wnt/betacatenin pathway. *Int J Mol Med*. 2020;45(5):1501-1513. doi: 10.3892/ijmm.2020.4525

38. Zhang WH, Yu LM, Han XX, Pan J, Deng JJ, Zhu LY, Liu YH. Human dental pulp stem cells conditioned medium protects genioglossus myoblast from cobalt chloride-induced hypoxia injury through AMPK/PGC-1α pathway. *Shanghai Journal of Stomatology*. 2020 Dec;29(6):573-579. doi: